\documentclass{elsart}

\usepackage{amssymb,graphicx,hyperref}

\newcommand{\FOR}{Fortran~95}
\newcommand{\taylor}{{\tt taylor}}

\newcounter{bla}
\newenvironment{refnummer}{%
\list{[\arabic{bla}]}%
{\usecounter{bla}%
 \setlength{\itemindent}{0pt}%
 \setlength{\topsep}{0pt}%
 \setlength{\itemsep}{0pt}%
 \setlength{\labelsep}{2pt}%
 \setlength{\listparindent}{0pt}%
 \settowidth{\labelwidth}{[9]}%
 \setlength{\leftmargin}{\labelwidth}%
 \addtolength{\leftmargin}{\labelsep}%
 \setlength{\rightmargin}{0pt}}}
 {\endlist}

\begin{document}


\journal{Computer Physics Communications}
\date{22 October 2009}

\begin{frontmatter}


\title{TaylUR 3,
  a multivariate arbitrary-order automatic differentiation package for \FOR}

\author{G.M. von Hippel\thanksref{a}}
\thanks[a]{Corresponding author}
\address{NIC, DESY, Platanenallee 6, 15738 Zeuthen, Germany}

\ead{georg.von.hippel@desy.de}
\ead[url]{http://www-zeuthen.desy.de/\~\/hippel}

\begin{abstract}
This new version of TaylUR is based on a completely new core,
which now is able to compute the numerical values of all of
a complex-valued function's partial derivatives up to an
arbitrary order, including mixed partial derivatives.
\end{abstract}

\begin{keyword}
automatic differentiation \sep higher derivatives \sep \FOR\/
\PACS 02.60.Jh \sep 02.30.Mv
\MSC 41-04 \sep 41A58 \sep 65D25
\\{\em Classification:} 4.12 Other Numerical Methods,
                      4.14 Utility
\end{keyword}

\end{frontmatter}


{\bf NEW VERSION PROGRAM SUMMARY}

\begin{small}
\noindent
{\em Program Title:} TaylUR \\
{\em Program Version:} 3.0 \\
{\em CPC Catalogue identifier:}
  \href{http://cpc.cs.qub.ac.uk/summaries/ADXR_v3_0.html}{ADXR\_v3\_0} \\
{\em Licensing provisions:} GPLv2 (see additional comments below) \\[2ex]
{\em No. of lines in distributed program:} 6750 \\
{\em No. of bytes in distributed program:} 19 162 \\
{\em Distribution format:} tar.gz \\[2ex]
{\em Programming language:} Fortran 95 \\
{\em Computer:}  Any computer with a conforming Fortran 95 compiler \\
{\em Operating system:} Any system with a conforming Fortran 95 compiler \\[2ex]
{\em CPC Catalogue identifier of previous version:} ADXR\_v2\_0 \\
{\em Journal reference of previous version:}
  Comput.~Phys.~Commun. {\bf 176} (2007) 710-711 \\

{\em Nature of problem:}\\
  Problems that require potentially high orders of partial
  derivatives with respect to several variables or derivatives
  of complex-valued functions, such as e.g. momentum or mass
  expansions of Feynman diagrams in perturbative QFT, and which
  previous versions of TaylUR [1,2] cannot handle due to their
  lack of support for mixed partial derivatives. \\
   \\
{\em Solution method:}\\
  Arithmetic operators and Fortran intrinsics are overloaded to act
  correctly on objects of a defined type {\tt taylor}, which encodes a
  function along with its first few partial derivatives with respect
  to the user-defined independent variables. Derivatives of products and
  composite functions are computed using multivariate forms [3] of
  Leibniz's rule
  \[
    D^{\nu}(fg) = \sum_{\mu \le \nu} 
                    \frac{\nu!}{\mu!(\mu-\nu)!} D^\mu f D^{\mu-\nu}g
  \]
  where
  $\nu=(\nu_1,\ldots,\nu_d)$,
  $|\nu|=\sum_{j=1}^{d}\nu_j $,
  $\nu!=\prod_{j=1}^{d}\nu_j!$,
  $D^{\nu} f =
    \frac{\partial^{|\nu|}f}{\partial^{\nu_1} x_1\cdots\partial^{\nu_d} x_d}$,
  and
  $\mu<\nu$ iff either $|\mu|<|\nu|$ or $|\mu|=|\nu|,\,\mu_1=\nu_1,\,\ldots
   \mu_k=\nu_k,\mu_{k+1}<\nu_{k+1}$ for some $k\in\{0,\ldots,d-1\}$,
  and of F\`aa di Bruno's formula
  \[
    D^{\nu} (f\circ g) = \sum_{p=1}^{|\nu|} (f^{(p)}\circ g)
                         \sum_{s=1}^{|\nu|}
                         \sum_{\{(k_1,\ldots,k_s;\lambda_1,\ldots,\lambda_s)\}}
                           \frac{\nu!}{\prod_{j=1}^{s} k_j!\lambda_j!^{k_j}}
                           \left(D^{\lambda_j}g\right)^{k_j}
  \]
  where the sum is over
  \begin{eqnarray*}
    \{(k_1,\ldots,k_s;\lambda_1,\ldots,\lambda_s)\in \mathbb{Z}^{s(1+d)} &:&
      k_i>0,\, \mathbf{0}<\lambda_1<\ldots<\lambda_s,\,\\
      && \sum_{i=1}^s k_i=p,\,\sum_{i=1}^s k_i \lambda_i=\nu\}.
  \end{eqnarray*}
  An indexed storage system is used to store the higher-order derivative
  tensors in a one-dimensional array. The relevant indices
  $(k_1,\ldots,k_s;\lambda_1,\ldots,\lambda_s)$ and the weights occuring
  in the sums in Leibniz's and F\`aa di Bruno's formula are precomputed
  at startup and stored in static arrays for later use.\\
   \\
{\em Reasons for the new version:}\\
  The earlier version lacked support for mixed partial derivatives,
  but a number of projects of interest required them. \\ 
   \\
{\em Summary of revisions:}\\
  The internal representation of a \taylor\/ object has changed to
  a one-dimensional array which contains the partial derivatives
  in ascending order, and in lexicographic order of the corresponding
  multiindex within the same order. The necessary mappings between
  multiindices and indices into the \taylor\/ objects' internal
  array are computed at startup. \\
  To support the change to a genuinely multivariate \taylor\/ type,
  the {\tt DERIVATIVE} function is now implemented via an interface
  that accepts both the older format {\tt derivative(f,mu,n)}=
  $\partial_\mu^n f$ and also a new format {\tt derivative(f,mu(:))}=
  $D^{\mu}f$ that allows access to mixed partial derivatives.
  Another related extension to the functionality of the module is the
  {\tt HESSIAN} function that returns the Hessian matrix of second
  derivatives of its argument. \\
  Since the calculation of all mixed partial derivatives can be very
  costly, and in many cases only some subset is actually needed,
  a masking facility has been added. Calling the subroutine
  {\tt DEACTIVATE\_DERIVATIVE} with a multiindex as an argument
  will deactivate the calculation of the partial derivative
  belonging to that multiindex, and of all partial derivatives
  it can feed into. Similarly, calling the subroutine
  {\tt ACTIVATE\_DERIVATIVE} will activate the calculation of
  the partial derivative belonging to its argument, and of all
  partial derivatives that can feed into it. \\
  Moreoever, it is possible to turn off the computation of mixed
  derivatives altogether by setting {\tt Diagonal\_taylors} to
  {\tt .TRUE.}. It should be noted that any change of
  {\tt Diagonal\_taylors} or {\tt Taylor\_order} invalidates all
  existing \taylor\/ objects. \\
  To aid the better integration of TaylUR into the {\sc HPSrc}
  library [4], routines {\tt SET\_DERIVATIVE} and
  {\tt SET\_ALL\_DERIVATIVES} are provided as a means of manually
  constructing a \taylor\/ object with given derivatives. \\
   \\
{\em Restrictions:}\\
  Memory and CPU time constraints may restrict the number of variables
  and Taylor expansion order that can be achieved. Loss of numerical
  accuracy due to cancellation may become an issue at very high
  orders. \\
   \\
{\em Unusual features:}\\
  These are the same as in previous versions, but are enumerated again
  here for clarity. \\
  The complex conjugation operation assumes all independent variables
  to be real. \\
  The functions {\tt REAL} and {\tt AIMAG} do \emph{not} convert to
  real type, but return an result of type \taylor\/
  (with the real/imaginary part of each derivative taken) instead.
  The user-defined functions {\tt VALUE}, {\tt REALVALUE} and
  {\tt IMAGVALUE}, which return the value of a \taylor\/ object as
  as a complex number, and the real and imaginary part of this
  value, respectively, as a real number are also provided. \\
  \FOR\/ intrinsics that are defined only for arguments of real type
  ({\tt ACOS}, {\tt AINT}, {\tt ANINT}, {\tt ASIN},
   {\tt ATAN}, {\tt ATAN2}, {\tt CEILING}, {\tt DIM},
   {\tt FLOOR}, {\tt INT}, {\tt LOG10}, {\tt MAX}, {\tt MAXLOC},
   {\tt MAXVAL}, {\tt MIN}, {\tt MINLOC}, {\tt MINVAL}, {\tt MOD},
   {\tt MODULO}, {\tt NINT}, {\tt SIGN})
  will silently take the real part of \taylor\/-valued arguments unless
  the module variable {\tt Real\_args\_warn} is set to {\tt .TRUE.}, in
  which case they will return a quiet NaN value (if supported by the
  compiler) when called with a taylor argument whose imaginary part
  exceeds the module variable {\tt Real\_args\_tol}. \\
  In those cases where the derivative of a function becomes undefined at
  certain points (as for {\tt ABS}, {\tt AINT}, {\tt ANINT}, {\tt MAX},
  {\tt MIN}, {\tt MOD}, and {\tt MODULO}), while the value is well defined,
  the derivative fields will be filled with quiet NaN values (if supported
  by the compiler). \\
   \\
{\em Additional comments:}\\
  This version of TaylUR is released under the second version of the 
  GNU General Public License (GPLv2). Therefore anyone is free to use
  or modify the code for their own calculations. As part of the licensing,
  it is requested that any publications including results from the use of
  TaylUR or any modification derived from it cite refs. [1,2] as well as
  this paper. Finally, users are also requested to communicate to the author
  details of such publications, as well as of any bugs found or of required
  or useful modifications made or desired by them. \\
  \\
{\em Running time:}\\
  The running time of TaylUR operations grows rapidly with both the number of
  variables and the Taylor expansion order. Judicious use of the
  masking facility to drop unneeded higher derivatives can lead
  to significant accelerations, as can activation of the
  {\tt Diagonal\_taylors} variable whenever mixed partial derivatives
  are not needed. \\
  \\
{\em Acknowledgments:}\\
  The author thanks Alistair Hart for helpful comments and suggestions.
  This work is supported by the Deutsche Forschungsgemeinschaft
  in the SFB/TR~09. \\
  \\
{\em References:}
\begin{refnummer}
\item
  G.~M. von~Hippel,
  TaylUR, an arbitrary-order diagonal automatic differentiation package
    for Fortran 95,
  Comput.~Phys.~Commun. {\bf 174} (2006) 569
  [physics/0506222].
\item
  G.~M. von~Hippel,
  New version announcement for TaylUR, an arbitrary-order diagonal
    automatic differentiation package for Fortran 95,
  Comput.~Phys.~Commun. {\bf 176} (2007) 710
  [arXiv:0704.0274].
\item
  G.~M. Constantine, T.~H. Savits,
  A multivariate Faa di Bruno formula with applications,
  Trans.~Amer.~Math.~Soc. {\bf 348} (1996) 2:503.
\item
  A. Hart, G.~M. von~Hippel, R.~R. Horgan, E.~H. M\"uller,
  Automated generation of lattice QCD Feynman rules,
  Comput.~Phys.~Commun. {\bf 180} (2009) 2698
  [arXiv:0904.0375].
\end{refnummer}

\end{small}

\end{document}